# Evidence of presolar SiC in the Allende *Curious Marie* calcium aluminum rich inclusion.

O. Pravdivtseva[1], F. L. H. Tissot[2], N. Dauphas[3], & S. Amari[1].

[1]Physics Department and McDonnell Center for the Space Sciences, Washington University, Saint Louis, MO 63130, USA

[2]The Isotoparium, Division of Geological and Planetary Sciences, California Institute of Technology, Pasadena, CA 91125, USA

[3]Origins Laboratory, Department of the Geophysical Sciences and Enrico Fermi Institute, The University of Chicago, Chicago, Il 60637, USA

**Calcium aluminum rich inclusions (CAIs) are one of the first solids to have condensed in the solar nebula, while presolar grains formed in various evolved stellar environments. It is generally accepted that CAIs formed close to the Sun at temperatures above 1500 K, where presolar grains could not survive, and were then transported to other regions of the nebula where the accretion of planetesimals took place. In this context, a commonly held view is that presolar grains are found solely in the fine-grained rims surrounding chondrules and in the low-temperature fine-grained matrix that binds the various meteoritic components together. Here we demonstrate, based on noble gas isotopic signatures, that presolar SiC have been incorporated into fine-grained CAIs in the Allende carbonaceous chondrite at the time of their formation, and have survived parent body processing. This finding provides new clues on the conditions in the nascent solar system at the condensation of first solids.**

Presolar grains acquired isotopic anomalies at formation and their presence in meteorites was first suggested based on their unique Ne[1] and Xe isotopic signatures[2], including Xe produced by the *s*-process nucleosynthesis[3]. Although it was known that the *s*-process (neutron capture on a slow time-scale) occurs in Asymptotic Red Giant Branch (AGB) stars, it took 10 years of painstaking work to isolate presolar SiC, the carrier of this Xe component, named Xe-G[4,5]. The most detailed noble gas study of presolar SiC was conducted on Murchison, resulting in isolation of the Murchison KJ and Murchison LQ 92–97 % pure SiC fractions[6,7]. Presolar SiC contains highly anomalous Si, C and N[8,9], as well as Kr-G and Ne-G noble gas components[10,5,7].

CAIs are one of the highest-temperature phases to condense from a hot solar gas upon cooling[11]. While many CAIs have experienced such severe reprocessing, including melting, that all textural evidence of condensation has been obliterated, fine-grained, irregularly-shaped CAIs appear to preserve their condensate origin to some degree[12]. The light noble gases, He, Ne and Ar, have been extensively studied in individual fine-grained and coarse-grained CAIs[13,14,15]. 22Ne excesses have been observed in CAIs from the CV carbonaceous chondrites and have been attributed to spallation reactions on Na[13,14,15], although a contribution from Ne-G carried by presolar SiC, was also suggested[16]. Spallation noble gases are produced by interactions of primary and secondary cosmic ray particles with target atoms, and thus are cosmogenic in origin. Xenon isotopic composition was measured in CAIs that were neutron-irradiated for the I-Xe dating[17,18]. Krypton and Xe systematics in unirradiated CAIs were only once cursory analyzed and discussed[15]. The large isotopic anomalies for other elements[19,20] in bulk CAIs lead to a conclusion that the isotopic fingerprint of CAIs cannot be reproduced by a single presolar phase[20], and the carriers of the reported isotopic signatures[19,20] remain unidentified. Although CAIs, the oldest dated solar system solids[21], have been extensively studied, questions still remain regarding the nature and origin of the isotopic anomalies that they carry, their distribution among primitive meteoritic classes, and the relationships to other meteoritic components[22].

The *Curious Marie* inclusion studied here is a fine-grained CAI that is characterized by group II Rare Earth Elements (REE) fractionation pattern[23], indicating condensation from the nebula after loss of an ultra-refractory component[24]. Initially, we observed small but statistically significant $^{130}$Xe enrichments, consistent with *s*-process contributions, in the course of our step-wise pyrolysis of the *Curious Marie* aliquot which was neutron-irradiated for I-Xe dating[25]. This isotopic anomaly pointed towards SiC, the carrier of *s*-process Xe-G, which is not expected to be found in CAIs[26]. To investigate the source of the observed $^{130}$Xe enrichment, we measured the isotopic compositions of Xe, Kr, Ar and Ne in an unirradiated *Curious Marie* sample following the same step-wise extraction experimental protocol. Noble gas analyses were conducted at Noble gas Laboratory (Physics Department of Washington University, Saint Louis, USA) using two built-in-house mass-spectrometers with high transmission Baur-Signer ion sources. The combination of high-sensitivity mass-spectrometry and high-resolution step-wise pyrolysis utilized in our study was crucial for disentangling the isotopic compositions of noble gases in *Curious Marie*. Details of the sample preparation and the analytical techniques are described in the Methods section.

**Xenon**

Xenon-G is characterized by strong enrichment in the even-numbered $^{128}$Xe, $^{130}$Xe and $^{132}$Xe over the other six stable Xe isotopes. It was initially observed[3] in a severely etched sample of the CM carbonaceous chondrite Murchison at a concentration of $5\times10^{-11}$ cm$^3$ STP/g, a small fraction of the total $13.41\times10^{-8}$ cm$^3$ STP/g $^{132}$Xe. That sample was etched to remove the carrier of the dominant trapped Xe component[27] so that the small isotopic anomalies could be detected. Fine-grained CAIs are known to be depleted in volatiles, including Xe, and the concentration of the trapped Xe component in Allende CAIs is typically $10^{-10}$ cm$^3$ STP/g of $^{132}$Xe[17,18], eliminating the need for a chemical treatment. These CAIs are therefore favorable samples to search for *s*-process enrichments.

Xenon in the *Curious Marie* CAI is dominated by radiogenic $^{129*}$Xe due to decay of now extinct $^{129}$I ($T_{1/2}$ = 16 Ma). In this sample, the radiogenic $^{129*}$Xe correlates with an excess of $^{128}$Xe produced by natural neutron capture from cosmic ray secondary neutrons on $^{127}$I (Figure 1a). Although a small *s*-process contribution to $^{128}$Xe is also expected if a SiC carrier is present, it cannot be resolved here since neither the parent nuclei $^{127}$I concentration nor the natural neutron flux and fluence are known, and the overwhelming correlation with $^{129*}$Xe shows that the cosmogenic component dominates.

Release profiles of $^{129*}$Xe and $^{128*}$Xe indicate the presence of a major iodine mineral carrier phase characterized by melting in the 1200–1300°C temperature range with 99.95% of $^{129}$Xe and $^{128}$Xe released below 1450°C. The concentration of $^{129*}$Xe is $3.5\times10^{-8}$ cm$^3$ STP/g, compared to $1.0\times10^{-10}$ cm$^3$ STP/g for trapped $^{132}$Xe, and is consistent with previous observations for fine-grained Allende CAIs[17,18,25]. There is also an even smaller contribution from $^{244}$Pu fission, $0.3\times10^{-10}$ cm$^3$ STP/g of $^{132}$Xe (Figure1b). Small excesses of $^{130}$Xe were observed and were not accompanied by excesses in $^{124,126,131}$Xe, which would be expected in case of spallation, suggestive of Xe-G and consistent with our earlier observations for the neutron irradiated *Curious Marie* aliquot[25].

Nucleosynthesis due to slow neutron capture contributes to $^{128,129,130,131,132}$Xe isotopes to varying degrees[7,28], while *s*-process contributions to $^{134,136}$Xe are essentially zero. Because of this, the corrections for the $^{244}$Pu fission Xe were based on the measured $^{134}$Xe/$^{136}$Xe ratios[29]. After the subtraction of these Pu-fission contributions, the Xe isotopic composition for $^{130-136}$Xe isotopes was a mixture of Xe-Q[27], Xe-G and cosmogenic Xe components (Figure 1c.). Decomposition was done for the two possible spallation compositions[30,31]. The concentration of $^{132}$Xe-G, obtained as an average of the two calculations, corresponded to ~ $4.8\times10^{-12}$ cm$^3$ STP $^{130}$Xe/g.

**Krypton**

All Kr isotopes are produced by the *s*-process, with $^{82,83,84}$Kr unaffected by flux and temperature-dependent branchings. Two Kr isotopes can be diagnostic of specific *s*-process nucleosynthesis conditions. 86Kr is produced by neutron capture on $^{85}$Kr at high neutron densities where capture dominates over decay. At low neutron fluxes, $^{85}$Kr ($t_{1/2}$ = 10.8 years) decays before it can capture another neutron. Krypton-80 is produced by neutron capture on $^{79}$Se; the abundance of which is sensitive to stellar temperatures. Thus, the $^{86}$Kr/$^{84}$Kr ratio is sensitive to the effective neutron density while the $^{80}$Kr/$^{84}$Kr ratio reflects the temperature of the stellar environment. It was previously suggested[10] that Kr-G originated from "stellar environments where the density of free neutrons was not the same". Later work[7] demonstrated that the $^{86}$Kr/$^{84}$Kr ratio increases with SiC grain size, implying that larger grains formed in higher neutron density environments than smaller ones.

Analyses of the *Curious Marie* CAI revealed $^{80*,82*}$Kr contributions from natural neutron-capture on $^{79,81}$Br, with a $^{82}$Kr/$^{80}$Kr ratio of 0.3855 ± 0.0014, consistent with the natural neutron-capture effects in Xe for this CAI and similar to previous observations[15] (Figure 2a). The concentration of trapped[27] $^{84}$Kr is 1.6×10$^{-10}$ cm$^3$ STP/g, comparable to the trapped Xe concentration. The concentration of Br-derived $^{82}$Kr is 3.5×10$^{-7}$ cm$^3$ STP/g. Based on the Kr-G/Xe-G ratio in Murchison KJ separates[7], the expected $^{84}$Kr-G concentration in *Curious Marie* is ~0.3×10$^{-11}$ cm$^3$ STP/g, about 3 times lower than calculated for $^{132}$Xe-G, making Kr less diagnostic here.

Nevertheless, the $^{86}$Kr/$^{84}$Kr ratio in *Curious Marie* shows slight excesses over the ($^{86}$Kr/$^{84}$Kr)$_Q$ value in 1250°, 1300° and 1350°C extractions (Figure 2b); the same temperature steps where Xe-G excesses are observed. The cosmogenic contribution to $^{86}$Kr is negligible[32], so the $^{86}$Kr enrichment at these temperatures is likely due to the *s*-process. Considering a Br-derived $^{82*}$Kr concentration of 3.5×10$^{-7}$ cm$^3$ STP/g, $^{80,82}$Kr-G in *Curious Marie* cannot be resolved from the dominant Br-derived component. Contributions from natural neutron-capture on $^{79,81}$Br are shown in the $^{82}$Kr/$^{84}$Kr *vs.* $^{80}$Kr/$^{84}$Kr three isotope plot (Figure 2a). The slope of the mixing line slightly decreases when 1250°, 1300° and 1350°C extractions are not considered in the correlation, supporting an *s*-process enrichment of $^{86}$Kr.

**Neon and Argon**

Ne-G is essentially pure $^{22}$Ne with theoretically predicted values for $^{20}$Ne/$^{22}$Ne and $^{21}$Ne/$^{22}$Ne of 0.0827 and 0.00059[7]. Theoretically predicted value for ($^{38}$Ar/$^{36}$Ar)$_G$ is 0.66 ± 0.20[7]. There are at least five distinct Ar components in *Curious Marie*: trapped, air, spallation, mono-isotopic radiogenic $^{40}$Ar and $^{36,38}$Ar produced *via* (n,β) reactions on chlorine. However, decomposition into the components is impossible since Ar has only three stable isotopes.

Neon data for *Curious Marie* are presented on the $^{21}$Ne/$^{22}$Ne *vs.* $^{20}$Ne/$^{22}$Ne plot (Figure 3a); with expanded data area shown in Figure 3b. The Ne composition is consistent with a mixture of Ne-G component and chondritic cosmogenic Ne. Other possible Ne endmembers are spallation on pure sodium Ne$_{sp}$-Na[33] and modeled isotopic composition[33] of Ne from sodalite, nepheline and melilite, mineral phases common in CAIs.

*Curious Marie* consists of widely-distributed powdery, porous aggregates of sodalite and nepheline, with interspersed regions of grossular and melilite. When Ne in CAIs is measured at melting in one bulk extraction, the resulting isotopic composition represents the "whole rock" data, an averaged value for all noble gas carrier phases present in CAI. While physical mineral disaggregation is not feasible here, the step-wise pyrolysis can potentially extract noble gases carried by different mineral phases based on their different thermal properties. This is illustrated by the release profiles of $^{21,22}$Ne (Figure 3c) and the $^{21}$Ne/$^{22}$Ne *vs.* $^{20}$Ne/$^{22}$Ne three isotope plot where the experimental data points do not fall on a single mixing line but rather group according to their release temperatures (Figure 3b). Both $^{21}$Ne and $^{22}$Ne are produced by spallation, interactions of primary and secondary cosmic ray particles with target atoms. Their production rates depend on the size of a meteoroid, and the location of the sample within, collectively called the shielding parameter, as well as the elemental composition of the sample. Since the shielding parameter for the mineral phases in *Curious Marie* is the same, the $^{21}$Ne/$^{22}$Ne ratios measured here depend only on the elemental composition and will be different for the different mineral phases.

The 800°C data point shows the highest $^{22}$Ne enrichment, consistent with melting of sodalite/nepheline, two major mineral phases in *Curious Marie*. Low temperature sodalite is expected to be completely molten at ~1080°C; overlapping release of its Ne with one from nepheline that melts at ~1150°C, depending on its composition. These two major mineral phases are present in *Curious Marie* in the form of powdery porous aggregates that would melt at a eutectic temperature that is lower than the melting points of nepheline and sodalite (Figure 3c). This is supported by a sharp release of $^{36}$Ar in the 800°–1000°C extraction step, resulting in $^{38}$Ar/$^{36}$Ar of 0.0199 ± 0.0004 (Supplementary Table 4). Natural thermal neutrons capture on $^{35,37}$Cl produces 3000 times more $^{36}$Ar than $^{38}$Ar, so an $^{38}$Ar/$^{36}$Ar ratio is expected to be below the $(^{38}$Ar/$^{36}$Ar$)_Q$ value of 0.19[27] at the melting of a Cl-rich sodalite.

The Ne extractions above melting of sodalite/nepheline aggregates are consistent with the release of cosmogenic Ne from melilite where Ca was partially replaced by Na (Figure 3c). The $^{21}$Ne/$^{22}$Ne isotopic ratios in the 800–1150°C temperature range result from a superposition of the sodalite/nepheline and melilite cosmogenic Ne. We assume the extraction points with the highest $^{21}$Ne/$^{22}$Ne ratios, observed at 1200°C and 1250°C, to be representative of the cosmogenic $^{21}$Ne/$^{22}$Ne value characteristic of the *Curious Marie* melilite. Above 1250°C all changes relative to the highest $^{21}$Ne/$^{22}$Ne ratio could be explained by a contribution from $^{22}$Ne-G from SiC. We can exclude spallation on Na as a source of these changes, since the concentration of Na in melilite is lower than in sodalite/nepheline. Grossular melts at 2727°C and would contribute to the cosmogenic Ne inventory starting from ~1360°C, at the onset of self-diffusion, and can only increase $^{21}$Ne/$^{22}$Ne values. The concentration of $^{22}$Ne-G can be then conservatively estimated as $0.1\times10^{-8}$ cm$^3$ STP/g in 1300°C – 1450°C temperature extractions, assuming a binary mixture of Ne-G and cosmogenic Ne of the *Curious Marie* melilite composition (Figure 3b).

For Ar, presence of spallation and Ar-G components in melilite will result in $^{38}$Ar/$^{36}$Ar contributions that will both move $^{38}$Ar/$^{36}$Ar value above the $(^{38}$Ar/$^{36}$Ar$)_Q$ of 0.19. Although $^{38}$Ar/$^{36}$Ar ratios in *Curious Marie* fall into the 0.31 ÷ 0.53 range in 1250°–1450°C extractions (Supplementary Table 4), indicating contribution from either spallation or *s*-process nucleosynthesis, or both, these contributions cannot be resolved.

The similar temperature range for release of Ne-G, Kr-G and Xe-G is a clear sign that those anomalies are carried by SiC. Due to the lack of convincing evidence for another carrier of Ne-G, Kr-G and Xe-G in primitive chondrites, in the following we assume that presolar SiC is the sole carrier of these noble gas components in *Curious Marie*. In the pure Murchison KJA SiC residue, 50% of Ne-G and 60% of Xe-G have been released in a single 1200°C – 1400°C extraction[7], consistent with what we observed for Xe in *Curious Marie*. Assuming similar release patterns for Ne and Xe here, the concentration of $^{22}$Ne-G in *Curious Marie* is ~0.14×10$^{-8}$ cm$^3$ STP/g.

The 1000°C, 1100°C and 1150°C extractions suggest a minor addition of solar Ne or/and trapped Ne-Q. Contribution from Ne-HL can be excluded because no Xe-HL, characteristic of meteoritic nanodiamond, was observed in the *Curious Marie* CAI. Addition of atmospheric Ne would be more pronounced at 800°C, but it is not seen here.

**The case for a presolar SiC carrier**

The $^{130}$Xe excesses are diagnostic of the Xe-G component in fine-grained CAIs, since *s*-process contributions on other Xe isotopes are masked by I-derived and fission Xe. Although in our I-Xe dating studies[17,25] we observed small and variable $^{130}$Xe excesses in some fine-grained Allende CV3 CAIs, including *Curious Marie*, these excesses could not be clearly attributed to *s*-process nucleosynthesis. The I-Xe technique involves irradiating samples with up to $2\times10^{19}$ thermal neutrons/cm$^2$ in order to convert $^{127}$I into $^{128}$Xe. Such high fluences may lead to enrichments on other Xe isotopes due to neutron capture on Te, Ba and the Xe isotopes themselves.

In this context, the contribution from the $^{129}$Xe(n,β)$^{130}$Xe reaction is particularly important. It is characterized by a capture cross section of 21 ± 7 σ, with a significant resonance integral of 250 ± 50 σ[34]. For most meteoritic samples, the contribution to $^{130}$Xe from neutron capture on $^{129}$Xe is negligible. However, it cannot be ignored for fine-grained CAIs, which are high in radiogenic $^{129*}$Xe but low in trapped Xe concentrations. Considering this and the neutron environment revealed here by neutron capture on I and Br, our results for the unirradiated *Curious Marie* aliquot demonstrate that less than 2 % of enrichment on $^{130}$Xe is due to neutron capture on $^{129*}$Xe and the rest can only be explained by *s*-process nucleosynthesis contributions.

The concentration of $^{132}$Xe-G in *Curious Marie* is ~10$^{-11}$ cm$^3$ STP/g. It is comparable to ~5×10$^{-11}$ cm$^3$ STP/g of $^{132}$Xe-G reported in a pioneering study of Murchison[3], indicating that the observed *s*-process enrichment is innate to the *Curious Marie* CAI and cannot be explained by Allende matrix contamination. *Curious Marie* was ~1.25 cm long and 0.75 cm wide, and great care was taken to avoid matrix contamination during separation of the sample for this study. The sample would have had to be almost pure matrix to yield the Xe-G concentration observed in this work, which is clearly not the case. The Ne isotopic composition in *Curious Marie* also rules out the possibility of a matrix contamination. It was demonstrated[14] that $^{20}$Ne/$^{22}$Ne always plots below ~0.9 for the interior portions of the fine-grained CAIs, consistent with a mixture of the cosmogenic Ne component and Ne-G and/or Ne$_{spa}$-Na. Admixture of matrix in other studies[13,15] increased the $^{20}$Ne/$^{22}$Ne ratio up to ~3.4. The highest $^{20}$Ne/$^{22}$Ne value in *Curious Marie* is 0.72 for the 1150°C temperature step.

Chemical separation of presolar grains from two Murchison aliquots in earlier studies[6,7] produced 92–97 % pure SiC fractions. Noble gas analyses of these fractions, further separated by size, revealed decrease of the Xe-G concentration with increasing grain size, consistent with low-energy ion implantation[6,7]. It was also demonstrated that the $^{22}$Ne/$^{130}$Xe and $^{86}$Kr/$^{82}$Kr ratios correlate in the G component[7]. Both ratios increase with average grain size of the Murchison SiC KJ separates (Figure 4). The ($^{86}$Kr/$^{82}$Kr)$_G$ vs. ($^{22}$Ne/$^{130}$Xe)$_G$ correlation is not defined for grains smaller than 0.38 µm, which have ($^{22}$Ne/$^{130}$Xe)$_G$ ≈ 680[7] (Figure 4.); and we cannot estimate the $^{86}$Kr/$^{82}$Kr ratio of the G-component in *Curious Marie* since the contribution from $^{81}$Br-derived $^{82*}$Kr cannot be corrected. But, based on the ($^{22}$Ne/$^{130}$Xe)$_G$ value of ~290 (Figure 4), the size distribution of SiC grains in *Curious Marie* is closer to the astronomically observed[35] range of 0.025–0.25 µm, and differs from the data reported for Murchison KJ separates. The estimation of the SiC abundance in *Curious Marie* relies heavily on the SiC size distribution that, in turn, depends on the ($^{22}$Ne/$^{130}$Xe)$_G$ value, defined here with high uncertainty. Assuming a SiC grain size of ~0.2 µm based on the ($^{22}$Ne/$^{130}$Xe)$_G$ value of ~290, the abundance of SiC in *Curious Marie* is ~7 ppm.

The upper limit for the abundance value for *Curious Marie* of ~25ppm is set by the $^{132}$Xe-G concentration in the Murchison KJA SiC separate[7] with mass-weighted mean grain diameter of 0.38µm. The lower limit for the abundance of SiC grains in bulk CV3 Allende was estimated earlier[36,37] as 0.0062 ± 0.0049 ppm based only on Ne-G concentration, about $10^3$ times lower than reported for other carbonaceous chondrites. Five CV3 chondrites have been examined in these previous studies[36,37] providing the bulk SiC abundance values that ranged from 0.0062 to 0.39 ppm. *In situ* search of presolar SiC in Insoluble Organic Matter (IOM) residues[38] yielded 5 ppm for the SiC abundance in RBT 04133, the only CV3 chondrite examined in ref. 38. Assuming the fine-grained Allende CAIs sampled SiC from the same reservoir as the Allende matrix, our data for *Curious Marie* suggest ~7 ppm of SiC for CV3 Allende, the abundance value that is consistent with the *in situ* studies[38] and higher than previous noble gas estimations[36,37]. Still, based on both the *Curious Marie* noble gas content and the *in situ* observations, the abundance of presolar SiC grains in CV3s falls below 11 – 47 ppm range reported for other primitive meteorites.

Assuming all of the chondrites formed from the homogeneous presolar dust reservoir, good correlation between the abundance patterns of presolar grains within a meteorite class and the petrologic type of the host meteorites was explained by the thermal and chemical processing on the host meteorites parent bodies[36,37]. Large variations of abundances and characteristics of presolar components across the classes were then attributed to the thermal processing in the solar nebula[37]. The estimated ($^{22}$Ne/$^{130}$Xe)$_G$ ≈ 290 for *Curious Marie* suggests that the low abundance of SiC in Allende could be, in part, due to a SiC size distribution in CV3 chondrites that is skewed towards smaller grain sizes compared to meteorites of other types. This can explain why the noble gas-derived CV3's SiC abundances are lower than what was observed in the *in situ* work. Both of these approaches to the studies of presolar materials will overlook very small grains, but loss of the fine-grained SiC would be more significant in chemically processed meteorites.

All Ne-G and ($^{22}$Ne/$^{130}$Xe)$_G$ based estimations of the presolar SiC abundances were done on meteorites that have been chemically treated to various degrees to remove trapped noble gas components and reveal less abundant exotic noble gas signatures unique to presolar grains. Longer harsh etching would preferentially destroy smaller, Xe-G rich, presolar SiC grains and erode the surfaces of the larger ones. Small grains would be also effectively lost during chemical separation where the size cutoff of presolar grains is determined by centrifugation. This is illustrated by the noble gas data for the minimally processed Murchison[39] that implied an at least 4 times higher abundance of SiC in Murchison bulk than inferred from the noble gas data[7] for Murchison KJ. Any chemical treatment will affect finer-grained SiC population in CV3 chondrites to a higher degree than a coarse-grained one in other chondrites. It is especially true for Allende that was progressively chemically etched in search for the presolar grains by a succession of researchers[36].

High-resolution transmission electron microscopy study of nano-diamonds isolated from acid dissolution residues of Murchison and Allende demonstrated the presence of ~1 SiC per 40 nano-diamonds[40] supporting the existence a sizable fine-grained SiC population in Allende. The observed SiC grains ranged from 0.0015 to 0.035 µm, about one order of magnitude smaller than the effective cutoff size for a typical presolar grain separation procedure. Considering the abundance of nano-diamonds in the bulk Allende of 340 ppm[36], at least 2 ppm of SiC grains were removed from the Allende inventory in a colloidal solution prior to the SiC abundance study that resulted in the $0.0062 \pm 0.0049$ ppm value[37].

**Implications**

Our step-wise isotopic analyses of Xe, Kr and Ne in *Curious Marie* reveal the presence of *s*-process nucleosynthetic products in this Allende CAI, and are supported by the *Curious Marie* Ar data. We observed similar Xe *s*-process enrichments in three of four previously analyzed fine-grained Allende CAIs[17], indicating that the carrier of *s*-process noble gases is not unique to *Curious Marie*. These isotopic signatures are most likely carried by presolar SiC, implying that SiC grains were present in the CAI forming region or in the place were condensate grains were agglomerated into fine-grained CAIs. The presolar oxides are also expected to be present in fine-grained CAIs, but their noble gas signature is unknown. Meteoritic nano-diamonds are the most abundant presolar material in meteorites[41], although their origin is still debated. Nevertheless, we did not observe Xe-HL, characteristic of nano-diamonds, in *Curious Marie*, indicating that these grains have been destroyed in a hot solar nebular environment. The experimentally determined kinetics of SiC volatilization as a function of temperature, gas composition and flow rate in gas mixtures, modeled to resemble a gas of solar composition, demonstrated that presolar SiC grains would survive short heating events associated with formation of CAIs[42], although the survival time depended on their size. It was also suggested that survival of presolar SiC grains for more than several thousand years in a hot ($\geq 900°C$) nebular gases was assisted by their being encased in a thin layer of minerals that were inert to reaction with solar nebular gas[42]. Allende nano-diamonds, on the other hand, when heated in vacuum, showed dramatic structural changes consistent with release of X-HL component starting from $900°C$[43].

The correlation of $^{86}Kr/^{82}Kr$ with grain size in pure SiC separates suggested that larger SiC grains acquired their Kr in a higher neutron density environment than smaller ones[7]. Thus *Curious Marie* sampled a distinct population of SiC, one that acquired its Kr at relatively low neutron density in the He shell of AGB star. If the fine-grained Allende CAIs are representative of refractory inclusions found in other meteorites, it may indicate that these CAIs condensed in a region of the solar nebula where the fine-grained SiC were more abundant than the coarse-grained ones. The mechanism behind this apparent heterogeneous distribution of SiC in the solar nebula is unclear. Size sorting of presolar SiC in the nebula is one possible explanation[16]; a late injection from a stellar source with low neutron density into a homogenized or partially homogenized presolar grain reservoir is also conceivable.

**References**

1. Black, D. C. & Pepin, R. O. Trapped neon in meteorites – II. *Earth & Planet, Sci. Letters* **6**, 395–405 (1969).

2. Lewis, R. S., Srinivasan, B. & Anders, E. Host phase of a strange xenon component in Allende. *Science* **190**, 1251–1262 (1975).

3. Srinivasan, B. & Anders, E. Noble gases in the Murchison meteorite. *Science* **201**, 51–56 (1978).

4. Tang, M. & Anders, E. Isotopic anomalies of Ne, Xe, and C in meteorites. II Interstellar diamond and SiC: Carriers of exotic noble gases. *Geochim. Cosmochim. Acta* **52**, 1235–1244 (1988).

5. Lewis, R. S., Amari, S, & Anders, E. Meteoritic silicon carbide: pristine material from carbon stars. *Nature* **348**, 293–298 (1990).

6. Amari, S., Lewis, R. S. & Anders, E. Interstellar grains in meteorites: I. Isolation of SiC, graphite, and diamond; size distributions of SiC and graphite. *Geochim. Cosmochim. Acta* **58**, 459–470 (1994).

7. Lewis, R. S., Amari, S. & Anders, E. Interstellar grains in meteorites: II. SiC and its noble gases. *Geochim. Cosmochim. Acta* **58**, 471–494 (1994).

8. Tang, M., Anders, E. Hoppe, P. & Zinner, E. Meteoritic SiC and its stellar sources: Implications for galactic chemical evolution. *Nature* **339**, 351–354 (1989).

9. Zinner, E., Tang, M. & Anders, E. Interstellar SiC in the Murchison and Murray meteorites: isotopic composition of Ne, Xe, Si, C, and N. *Geochim. Cosmochim. Acta* **53**, 3273–3290 (1989).

10. Ott, U., Begemann, F., Yang, J. & Epstein, S. (1988) *S*-process krypton of variable isotopic composition in the Murchison meteorite. *Nature* **332**, 700–702 (1988).

11. Grossman, L. Condensation in the primitive solar nebula. *Geochim. Cosmochim. Acta* **36**, 597–619 (1972).

12. Krot, A.N., Petaev, M.I., Russell, S.S., Itoh, S., Fagan, T.J., Yurimoto, H., Chizmadia, L., Weisberg, M.K., Komatsu, M., Ulyanov, A.A., Keil, K. *Chemie der Erde-Geochemistry* **64**, 185–239 (2004).

13. Smith, P. S., Huneke J. C., Rajan, R. S. & Wasserburg G. J. Neon and argon in the Allende meteorite. *Geochim. Cosmochim. Acta* **41**, 627–647 (1977).

14. Vogel, N., Baur, H., Bischoff, A., Leya, I. & Wieler, R. Noble gas studies in CAIs from CV3 chondrites: No evidence for primordial noble gases. *Meteoritics & Planetary Science* **39**, 767–778 (2004).

15. Göbel, R., Begemann F. & Ott U. On neutron induced and other noble gases in Allende inclusions. *Geochim. Cosmochim. Acta* **46**, 1777–1792 (1982).

16. Russel, S. S., Franchi, I. A., Verchovsky, A. B., Ash, R. D. & Pillinger, C. T. Carbon, nitrogen, and noble gases in Vigarano (CV) calcium-aluminum-rich inclusion: Evidence for silicon carbide in refractory inclusions (abstract). *Meteoritics & Planetary Science* **33**, A132 (1998).

17. Pravdivtseva, O. V, Krot, A. N., Hohenberg, C. M., Meshik, A. P., Weisberg, M. K. & Keil, K. The I-Xe record of alteration in the Allende CV chondrite. *Geochim. Cosmochim. Acta* **67**, 5011–5026 (2003).

18. Swindle, T. D., Caffee, M. W., Hohenberg, C. M. & Lindstrom, M. M. I-Xe studies of Allende inclusions: EGGs and the Pink Angel. *Geochim. Cosmochim. Acta* **47**, 2157–2177 (1988).

19. Burkhardt, C., Kleine, T., Oberli, F., Pack, A., Bourdon, B. & Wieler, R. Molybdenum isotope anomalies in meteorites: Constraints on solar nebula evolution and origin of the Earth. *Earth and Planetary Sci. Lett.* **312**, 390–400 (2011).

20. Shollenberger, Q.R., Render, J., & Brennecka, G.A. Er, Yb, and Hf isotopic compositions of refractory inclusions: An integrated isotopic fingerprint of the Solar System's earliest reservoir. *Earth and Planetary Sci. Lett.* **495**, 12–23 (2018).

21. Connelly, J. N., Bizzarro, M., Krot, A. N., Nordlund, Å., Wielandt, D., & Ivanova, M. A. The absolute chronology and thermal processing of solids in the solar protoplanetary disk. *Science* **338**, 651–655 (2012).

22. MacPherson, G. J., Simon, S.B., Davis, A. M., Grossman, L, & Krot, A. N. Calcium-aluminum-rich inclusions: Major unanswered questions. In: *Chondrites and the protoplanetary disk*. Astronomical Society of the Pacific Conference Series **341** (Editors: Krot, A. N., Scott, E. R. D. & Reipurth, B.) Sheridan Books, 225–247 (2005).

23. Tissot, F. L. H., Dauphas, N. & Grossman, L. Origin of uranium isotope variations in early solar nebula condensates. *Science Advances* **2**, (2016).

24. Boynton, W. V. Fractionation in the solar nebula: condensation of yttrium and rare earth elements. *Geochim. Cosmochim. Acta* **39**, 569–584 (1975).

25. Pravdivtseva, O., Meshik, A., Tissot F. L. H. & Dauphas, N. I-Xe studies of aqueous alteration in the Allende CAI *Curious Marie*. *Lunar Planet. Sci. XLIX*, #2959 (2018).

26. MacPherson, G. J. & Grossman, L. "Fluffy" type A Ca-Al-rich inclusions in the Allende meteorite. *Geochim. Cosmochim. Acta* **48**, 29–46 (1984).

27. Busemann, G., Baur, H., &Wieler, R. Primordial noble gases in "phase Q" in carbonaceous and ordinary chondrites studied by closed-system stepped etching. *Meteoritics & Planetary Science* **35**, 949–973 (2000).

28. Gallino, R., Busso, M., Picchio, G. & Raiteri, C.M. On the astrophysical interpretation of isotope anomalies in meteoritic SiC grains. *Nature* **348**, 298–302 (1990).

29. Lewis, R. S. Rare gases in separated whitlockite from the St. Severin chondrite: Xenon and krypton from fission of extinct $^{244}$Pu. *Geochim. Cosmochim. Acta* **39**, 417–432 (1975).

30. Hohenberg, C. M., Hudson, B., Kennedy, B. M. & Podosek F. A. Xenon spallation systematics in Angra dos Reis. *Geochim. Cosmochim. Acta* **45**, 1909–1915 (1981).

31. Kim, J. S. & Marti, K. Solar type xenon: isotopic abundances in Pesyanoe. *Proc. Lunar Planet Sci.* **22**, 145–151 (1992).

32. Lavielle, B., & Marti, K. Cosmic ray produced Kr in St. Severin core AIII. Proceedings of 18$^{th}$ Lunar and Planetary conference, 565–572 (1988).

33. Leya, I., Lange, H. J., Neumann, S., Wieler, R., & Michel, R. The production of cosmogenic nuclides in stony meteoroids by galactic cosmic ray particles. *Meteoritics & Planetary Science* **35**, 259–289 (2000).

34. Mughabghab, S. F. Thermal neutron capture cross-sections resonance integrals and g-factors. INDC(NDS)-440, IAEA, 33pp. (2003).

35. Mathis, J. S., Rumpl, W. & Nordsieck, K. H. The size distribution of interstellar grains. *The Astrophysical Journal* **217**, 425–433 (1977).

36. Huss, G. R. & Lewis, R. S. Presolar diamond, SiC and graphite in primitive chondrites: Abundances as function of meteorite class and petrologic type. *Geochim. Cosmochim. Acta* **59**, 115–160 (1995).

37. Huss, G. R., Meshik, A. P., Smith, J. B. & Hohenberg, C. M. Presolar diamond, silicon carbide, and graphite in carbonaceous chondrites: Implications for thermal processing in the solar nebula. *Geochim. Cosmochim. Acta* **67**, 4823–4848 (2003).

38. Davidson, J., Busemann, H., Nittler, L. R., Alexander, C., M., O'D., Orthouth-Daunay, F-R., Franchi, I. A. & Hoppe, P. Abundanses of presolar silicon carbide grains in primitive meteorites determined by NanoSIMS. *Geochim. Cosmochim. Acta* **139**, 248–266 (2014).

39. Ott, U. & Merchel, S. Noble gases and not so unusual size of presolar SiC in Murchison. *31$^{st}$ LPSC*, Abstract 1356.

40. Daulton, T. L., Eisenhour, D. D., Bernatowicz, T. J., Lewis, R. S. & Buseck, P. P. Genesis of presolar diamonds: Comparative high-resolution transmission electron microscopy study of meteoritic and terrestrial nano-diamonds. *Geochim. Cosmochim. Acta* **60**, 4853–4872 (1996).

41. Lewis, R. S., Anders, E. & Drain, B. T. Properties, detectability and origin of interstellar diamonds in meteorites. *Nature* **339**, 117–121 (1989).

42. Mendybaev, R. A., Beckett, J. R., Grossman, L., Stolper, E., Cooper, R. F., & Bradley, J. P. Volatilization kinetics of silicon carbide in reducing gases: An experimental study with applications to survival of presolar grains in the solar nebula. *Geochim. Cosmochim. Acta* **66**, 661–682 (2002).

43. Stroud, R. Structural and elemental transformation of meteoritic nanodiamonds during in situ heating in a UHV scanning transmission electron microscope. *82nd Annual Meeting of The Meteoritical Society,* Abstract #6457 (2019).

**Corresponding author** – Olga Pravdivtseva olga@physics.wustl.edu

**Acknowledgments** – We are grateful to D. Nakashima and two anonymous reviewers for the constructive and thorough reviews and to the help provided during the editorial process. This work was supported by NASA grant 80NSSC19K0508 to O.P.; a Crosby Postdoctoral Fellowships (MIT), NSF-EAR grant 1824002, and start-up funds provided by Caltech to F.T.; by NASA grants NNX17AE86G (LARS), NNX17AE87G 531 (Emerging Worlds), and 80NSSC17K0744 (Habitable Worlds) to N. D.

**Contributions** - O.P designed the study, conducted the noble gas analyses, treated the data and wrote the first draft of the paper. F. T. selected, prepared and characterized the studied sample. O.P. and S.A interpreted the data. N.D critically contributed to the paper presentation. All the authors contributed to the discussion of the results and commented on the manuscript.

**Figure Legends:**

**Figure 1.** Three-isotope plots showing Xe composition in the Allende *Curious Marie* CAI. Green diamonds represent various Xe components, red numbers are extraction temperatures, error bars are 1σ. (a) – the linear correlation between $^{128}$Xe and $^{129}$Xe (dashed line) is characteristic of iodine-derived Xe, where the slope of the line depends on the thermal neutron fluence. $^{129*}$Xe is a decay product of now extinct $^{129}$I ($T_{1/2}$ = 16 Ma). $^{128*}$Xe is produced by cosmic ray secondary neutron capture on $^{127}$I. (b) – contribution from $^{244}$Pu fission; addition of $^{132}$Xe-G shifts experimental points along the dashed correlation line. (c) – the Xe isotopic composition after subtraction of the $^{244}$Pu contribution is a three-component mixture of trapped Xe-Q, Xe-G and cosmogenic Xe, produced by cosmic ray particle spallation. Shown are two Xe cosmogenic compositions: Xe-cos1 – estimated from the Angra dos Reis angrite mineral fractions[30] and Xe-cos2 – calculated for the aubrite Pesyanoe[31]. The 1250°C experimental point shows the highest Xe-G contribution.

**Figure 2.** Krypton in the Allende *Curious Marie* CAI. (a) – Three-isotope plot demonstrating linear correlation between $^{82}$Kr and $^{80}$Kr that is characteristic of Br-derived Kr. $^{80*}$Kr and $^{82*}$Kr are produced by cosmic ray secondary neutron capture on $^{79,81}$Br at depth, where high energy cosmic ray particles have slowed down to the thermal and epithermal energies. Green diamonds represent various Kr components, red numbers are extraction temperatures, error bars are 1σ. (b) – $^{86}$Kr/$^{84}$Kr values at different extraction temperatures. Horizontal line corresponds to ($^{86}$Kr/$^{84}$Kr)-Q; small $^{86}$Kr excesses over Kr-Q in 1250°C, 1300°C and 1350°C extractions are consistent with *s*-process nucleosynthesis enrichments.

**Figure 3.** Neon in the *Allende Curious* Marie CAI. (a) – Three-isotope plot showing Ne data that are consistent with mixture of cosmogenic Ne, Ne-G and Ne$_{sp}$-Na. Green diamonds represent various Ne components, error bars are 1σ. Abbreviations stand for cosmogenic Ne modeled for targets of different mineral composition[33]: sod – sodalite; mel – melilite; neph – nepheline. (b) – Expanded data point area. Red numbers are extraction temperatures; outlined green areas correspond to the green diamonds in panel (a). All changes relative to the highest $^{21}$Ne/$^{22}$Ne ratio above 1200°C could be explained by a contribution from $^{22}$Ne-G from SiC. (c) – Release

profiles of $^{21,22}$Ne normalized to temperature step ΔT. Two well-defined peaks correspond to melting of sodalite/nepheline and melilite.

**Figure 4.** Correlation between $^{22}$Ne/$^{130}$Xe and $^{86}$Kr/$^{82}$Kr in the G noble gas component carried by the presolar SiC grains. Error bars are 1σ. Both ratios increase with average grain size of the Murchison KJ separates[6,7]. The vertical lines correspond to the ($^{22}$Ne/$^{130}$Xe)$_G$ ratios in minimally processed Murchison[39] (black) and in the Allende *Curious Marie* CAI (red).

## Methods

### Sample description

The fine-grained ~1.25×0.75 cm *Curious Marie* CAI (Supplementary Figure 1) was provided by the Robert A. Pritzker Center for Meteoritics and Polar Studies at the Chicago Field Museum of Natural History. The mineralogical study of the sample was conducted at the Origins Lab at the University of Chicago and is described in detail in ref. 23. For that work a small interior chip of *Curious Marie* was mounted in epoxy (Buehler). This sample consisted of widely-distributed powdery, porous aggregates of sodalite [Na$_8$(Al$_6$Si$_6$O$_{24}$)Cl$_2$], nepheline [(Na$_3$K)Al$_4$Si$_4$O$_{16}$], with interspersed regions of grossular (Ca$_3$Al$_2$Si$_3$O$_{12}$), and melilite ((Ca,Na)$_2$(Al,Mg,Fe$^{2+}$)[(Al,Si)SiO$_7$]), indicating high degree of aqueous alteration involving Na-Cl-bearing fluids[44]. Inductively Coupled Plasma Mass Spectrometry (ICP-MS) measurements demonstrated that *Curious Marie* CAI is characterized by a Group II rare earth element (REE) pattern[23,45], consistent with a condensation origin of its precursor minerals[46]. Although sodalite is typically present in the peripheral areas of Allende CAIs[47,48], it is widely distributed throughout *Curious Marie*.

Nucleosynthetic anomalies in the relatively large, centimeter-sized meteoritic components are usually very limited. Nevertheless, some CAIs with Fractionation and Unknown Nuclear effects (FUN CAIs) display very large isotopic anomalies and mass-dependent isotopic fractionation. These FUN CAIs are either compact Type A or coarse-grained Type B CAIs. They are recognized by higher than ~+10 ‰ δ$^{25}$Mg mass fractionation effects and/or a deficit in δ$^{26}$Mg*, ~ 10 ‰/u Ti mass-dependent effects and large $^{50}$Ti nucleosynthetic anomalies. Isotopic composition of oxygen in FUN CAIs does not follow the CAIs mass fractionation line. *Curious Marie* does not demonstrate any of the FUN CAIs characteristics. Its oxygen isotopic composition falls on the CAIs mass fractionation line [49]; it has minor $^{50}$Ti anomaly[50] and only ~ -0.28 ‰/u Ti mass-dependent fractionation[50]. The $^{26}$Mg excess in Curious Marie is resolvable and the δ $^{25}$Mg values range from -9 to -19 ‰[50]. While the negative δ $^{25}$Mg observed in *Curious Marie* seems to be a common characteristic of fine-grained CAIs[51,52], the δ $^{25}$Mg values in coarse-grained CAIs range from about +5 to +10‰, and FUN CAIs have δ $^{25}$Mg values that are higher than +10 ‰[51,52]. Thus, it was concluded, based on the petrology data and oxygen, Mg and Ti isotopic composition, that *Curious Marie* is not a FUN CAI[23,50].

### Mass-spectrometry

20.4 mg of *Curious Marie* were extracted from Allende chip using cleaned stainless steel dental tools, wrapped in 0.005" thick Pt foil and placed under vacuum into the sample system of mass-spectrometer at Noble gas Laboratory (Physics Department of Washington University, Saint Louis, USA). The sample was kept at 150°C for two days to remove surface-bound gases and transferred under vacuum (without exposing it to the atmosphere) into the resistance furnace for pyrolysis. Noble gases were released by step-wise extractions in a low-blank W-coil starting with 800°C and up to the melting of Pt (1770°C), with 50°C steps in 1100–1700°C interval. The extraction temperatures during the pyrolysis were determined indirectly *via* the heating coil current. The heating coil temperature as a function of heating coil current was calibrated using an optical pyrometer and W-WRe thermocouple. The double-walled radiation Ta-shield around the heating coil insures good agreement between actual extraction temperatures within the coil and the coil temperatures established by the calibration. Since the samples are routinely wrapped in platinum foil for the analyses, the heating coil current, recorded for each studied sample at the melting of platinum, provides an independent internal calibration.

Two built-in-house mass-spectrometers[53] with high transmission Baur-Signer ion sources were used for the analyses. They shared the extraction and sample purification system. Released noble gases were purified sequentially by first exposing them to two SAES St707 pellet getters, maintained at 275°C, and then to 3 freshly deposited Ti–film getters. Cleaned noble gases were split between SuperGnome N, optimized for the measurements of Xe and Kr, and SuperGnome S optimized for Ne and Ar . Xenon, Kr and Ar were frozen on activated charcoal at -196°C with Ne isotopic composition being measured first. Then, temperature on charcoal was increased to 125°C, effectively releasing Ar that was analyzed next. Finally, a charcoal temperature was elevated to 165°C for Xe and Kr desorption, the isotopic compositions of these two heavy noble gases were analyzed together. The sensitivity for $^{132}$Xe was $6.21\times10^{-16}$ cm$^3$ STP/count, for $^{84}$Kr – $6.37\times10^{-16}$ cm$^3$ STP/count, and for $^{20}$Ne – $1.03\times10^{-14}$ cm$^3$ STP/count.

Isotopic compositions of noble gases and the peaks at m/e = 18, 19, 23, 40, and 44, required for the corrections of the Ne interferences, were measured in a peak-jumping mode. Twenty five mass-jumping cycles were used to measure Kr and Xe and twenty to measure Ne and Ar. Count rate for each mass peak was averaged from 5 to 25 1-second integrations, to optimize counting statistics. Isotopic ratios were corrected for isobaric interferences at each cycle, extrapolated to time "zero" (when the source high voltage was turned on) and then determined using the least-squares fitting program (Supplementary Tables 1–4).

Procedural blanks were measured with an empty W-coil at 1200°C, 1500°C (15 min). An additional "hot" blank was measured after the completion of the step-wise heating analysis, at ~ 1780°C, above the melting of Pt. Procedural blanks were consistent with each other and for Kr and Xe were negligible, typically at $2\times10^{-15}$ $^{132}$Xe cm$^3$ STP. Ne data were corrected for blank (see **Data reduction**).

In both mass-spectrometers the electron multipliers with Be-Cu dynodes and conversion electrodes were employed. Discrimination level was determined from pulse height distribution plot and it was found to be different for light and heavy noble gases. We use two-channel ion counters (SR400). During the measurements these two channels were set to two different discrimination levels, just 3 mV apart. At the end of each measurement the cumulative count rates for two channels were compared to ensure the correct discrimination during the analyses. While this approach does not improve counting statistics, it serves as a sensitive internal check of proper vacuum condition and performance of counting system during the analysis,

**Data reduction**

We treat raw data by various processing programs. First, the interference corrections were calculated and applied on a cycle-by-cycle basis, then the instrumental mass discrimination and dead-time corrections were made, and finally a file was generated that keep track of all correlated errors. Simple error propagation cannot be applied when isotopic ratios are involved and components are subtracted (blanks, corrections, *etc.*). Correlated errors were introduced since the isotopic ratios are not truly independent, and correct error propagation involves computations with an error tensor, a N×N array, where N is the number of isotopes.

Xenon and Kr data reductions were trivial, only small 0.1%/u (typical to the Baur Singer ion source) and reproducible instrumental mass discrimination corrections were applied.

During Ne measurements masses 18 ($H_2O$), 19 (F), 40 ($^{40}Ar$) and 44 ($CO_2$) were measured in order to account for the $H_2O^+$, $HF^+$, $^{40}Ar^{++}$ and $CO_2^{++}$ interferences. Additionally, mass 23 was measured to estimate contribution of $^{20}NeH^+$ to $^{21}Ne^+$. Twenty sweeps of these species plus three Ne isotopes were used to determine the Ne composition of the sample. For each cycle, similar corrections were performed for the signal at mass 20 based on the interferences from $H_2O^+$ ($H_2^{18}O^+$, $DH^{17}O^+$, $D_2^{16}O^+$) and from $HF^+$ using the count rates at masses 18 and 19 respectively. Corrections for the signal at mass 22 were applied based on interference from $CO_2^{++}$ using the count rates at mass 44.

To determine the associated interference correction factors, residual gas with various proportions of F, $^{40}Ar$ and $CO_2$ was analyzed. The count rates at masses 20 and 22 were used to determine the factors $HF^+/F^+$, $^{40}Ar^{++}/^{40}Ar^+$ and $CO_2^{++}/CO_2^+$ (0.0084±0.0008; 0.131±0.003; 0.0063±0.0006, respectively). The mass 18 ($H_2O$) correction factor (~0.0025= signal at mass 20 due to $H_2O$ divided by signal at mass 18) is based on the terrestrial isotopic abundances of H and O.

Neon measurements were performed at electron energy 48 eV slightly above the threshold for double ionization of Ar. In addition, during the Ne analyses the background Ar was kept low by cold (-196°C) charcoal finger close to the ion source.

**Contribution from the $^{129}Xe(n,\beta)^{130}Xe$ reaction**

The slope of the $^{129}Xe/^{132}Xe$ versus $^{128}Xe/^{132}Xe$ isochron for the MURR irradiated sample of *Curious Marie* is 1200 times less than for the naturally irradiated Curious Marie aliquot. This means *Curious Marie* experienced a thermal equivalent cosmogenic neutron fluence of $1.6\times10^{16}$ n/cm$^2$. Considering the concentration of $^{129}Xe$ of $3.5\times10^{-8}$ cm$^3$ STP/g in *Curious Marie*, the concentration of neutron-induced $^{130}Xe$ is $1.4\times10^{-13}$ cm$^3$ STP/g, 34 times less than the excess attributed to $^{130}Xe$-G. Thus only ~2% of estimated $^{130}Xe$-G could be due to neutron capture on $^{129}Xe$. The major uncertainties here are the unknown neutron spectrum for MURR (Missouri University Research Reactor) and the energy spectrum of the "natural" neutrons.

**Ne data**

*Curious Marie* consists of widely-distributed powdery, porous aggregates of sodalite [$Na_8(Al_6Si_6O_{24})Cl_2$] and nepheline [$(Na_3K)Al_4Si_4O_{16}$], with interspersed regions of grossular ($Ca_3Al_2Si_3O_{12}$) and melilite (($Ca,Na)_2(Al,Mg,Fe^{2+})[(Al,Si)SiO_7]$).

The calculations of isotopic compositions of cosmogenic Ne from mineral phases, outlined by the green boxes in Figure 3b, are based on the simplified chemical compositions and are shown for reference. The area for melilite is based on the $Ca_2Al_2SiO_7$ gehlenite end member[54]. Two end members of the melilite series, Åkermanite and gehlenite melt congruently at 1451°C and 1590°C. They form a solid solution with a minimum melting temperature of 1385°C at the composition $Åk_{72}Ge_{28}$. Natural melilites, like one in the *Curious Marie* CAI, contain appreciable amounts of iron and sodium. The replacement of (Ca,Mg,Al) by (Na,$Fe^{2+}$,$Fe^{3+}$) causes a marked lowering of the melting temperature, which is reflected here in the melilite $^{21,22}$Ne release peak (Figure 3c).

Since the concentration of Na in melilite is lower than in sodalite/nepheline, we assume the extraction points with the highest $^{21}$Ne/$^{22}$Ne ratios, observed at 1200°C and 1250°C, to be representative of the cosmogenic $^{21}$Ne/$^{22}$Ne value characteristic of the *Curious Marie* melilite.

**Data availability**

The data that support the plots within this paper and other findings of this study are provided in the Supplementary Information section as Supplementary Tables 1–4; and are also available from the corresponding author upon reasonable request.

**References**

44. Kimura, M. & Ikeda, Yu. Anhydrous alteration of Allende chondrules in the solar nebula II: Alkali-Ca exchange reactions and formation of nepheline, sodalite and Ca-rich phases in chondrules. *Proc. NIPR Simp. Antarct. Meteorites* **8**, 123–138 (1995).

45. Boynton, W. V. Fractionation in the solar nebula: condensation of yttrium and rare earth elements. *Geochim. Cosmochim. Acta* **39**, 569–584 (1975).

46. Davis, A. M. & Grossman, L. Condensation and fractionation of rare earth in the solar nebula. *Geochim. Cosmochim. Acta* **43**, 1611–1632 (1979).

47. Krot, A. N., Petaev, M. I. & Bland, P. A. Multiple formation mechanisms of ferrous olivine in CV carbonaceous chondrites during fluid-assisted metamorphism. *Antarctic Meteor. Res.* **17**, 153–171 (2004).

48. Krot, A. N., Nagashima, K., Hutcheon, I. D., Ishii, H. A., Yin, Q.-Z., Davis, A. M. & Simon, S. B. Mineralogy, petrography, oxygen and magnesium isotopic compositions and formation age of grossular-bearing assemblages in the Allende CAIs. *Lunar Planet. Sci Conf.* 41, Abstract #1406 (2010).

49. Brearley, A. J. & Krot, A. N. Metasomatism in chondritic meteorites. In: *Metasomatism and the Chemical Transformation of Rock: The Role of Fluids in Terrestrial and Extraterrestrial Processes*. Lecture Notes in Earth Sciences Series. (Editors: Dan Harlov and Håkon Austrheim) Springer Verlag, 653–782 (2013).

50. Tang, H., Liu, M.-Ch., McKeegan, K. D., Tissot, F. L. H. & Dauphas, N. *In situ* isotopic studies of the U-depleted Allende CAI *Curious Marie*: Pre-accretionary alteration and the co-existence of $^{26}$Al and $^{36}$Cl in the early solar nebula. *Geochim. Cosmochim. Acta* **207**, 1–18 (2017).

51. Krot, A. N., Scott, E. R. & Zolensky, M. E. Origin of fayalitic olivine rims and lath-shaped matrix olivine in the CV3 chondrite Allende and its dark inclusions. *Meteor. Planet. Sci.* **32**, 31–49 (1997).

52. Krot, A. N., Nagashima, K., Hutcheon, I. D., Ishii, H. A., Jacobsen, B., Yin, Q.Z., Davis, A. M. & Simon, S. B. Mineralogy, petrography, oxygen and magnesium isotopic compositions and formation age of grossular-bearing assemblages in the Allende CAIs. *Lunar and Planetary Science Conference XLI*, Abstract#1406 (2010).

53. Hohenberg, C. M. High sensitivity pulse-counting mass-spectrometer system for noble gas analysis. *Rev. Sci. Instrum.* **51** 1075–1082 (1980).

54. Deer, W. A., FRS, Howie, R. A. & Zussman, J. *An Introduction to the Rock-Forming Minerals.* (Mineralogical Society of Great Britain and Ireland, 2013).

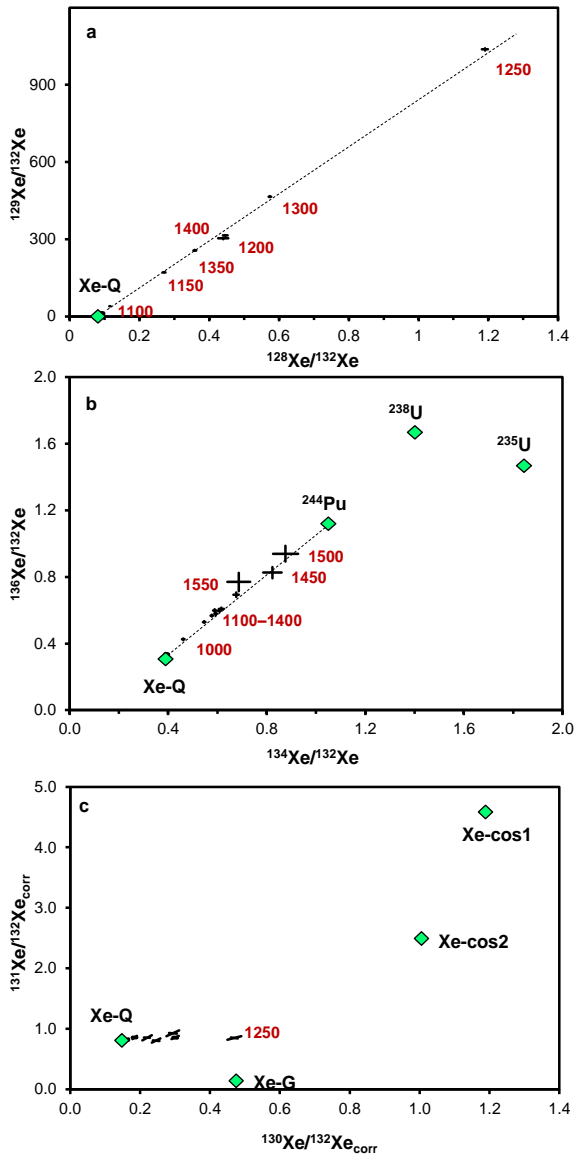

Figure 1

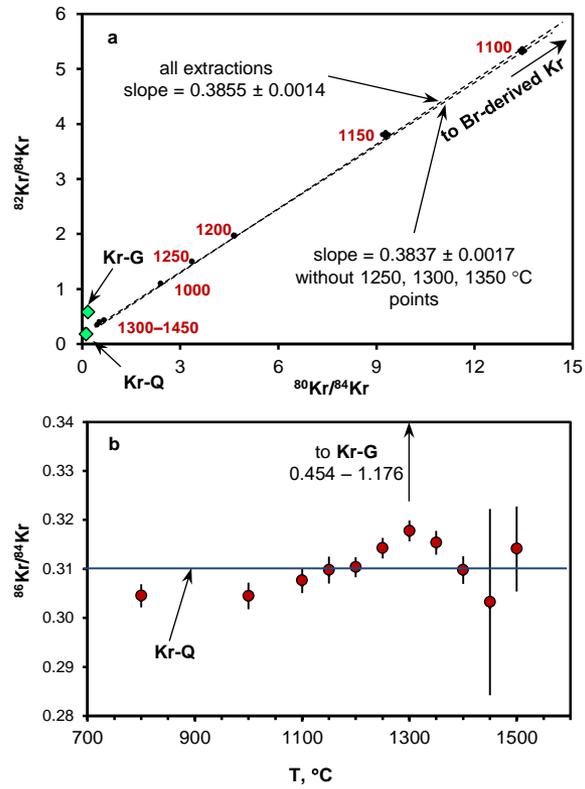

Figure 2

Figure 3

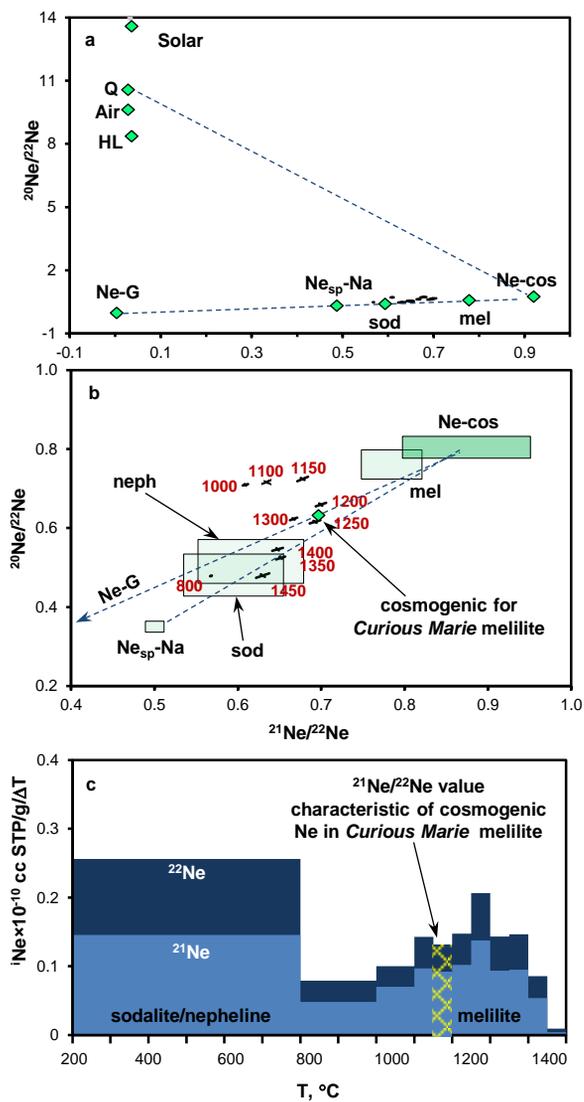

Figure 4

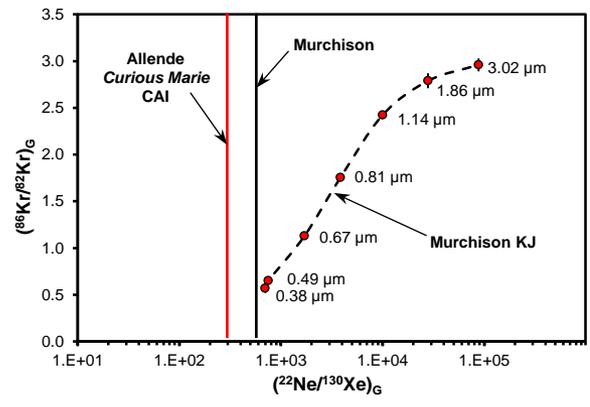